\documentclass[a4paper]{jpconf}
\usepackage{graphicx}

\usepackage[square, sort&compress, comma, numbers]{natbib}

\begin{document}
\title{Multi-messenger astronomy with very-high-energy gamma-ray observations}

\author{Jim Hinton$^{1,*}$ \& Edna Ruiz-Velasco$^{1,**}$}

\address{$^{1}$Max-Planck-Institut für Kernphysik, P.O. Box 103980, D 69029 Heidelberg, Germany}

\ead{$^{*}$hinton@mpi-hd.mpg.de, $^{**}$edna.ruiz@mpi-hd.mpg.de}

\begin{abstract}
After decades of development, multi-messenger astronomy, the combination of information on cosmic sources from photons, neutrinos, charged particles and gravitational waves, is now an established reality. Within this emerging discipline we argue that very-high-energy (VHE) gamma-ray observations play a special role.  We discuss the recent progress on explosive transients, the connections between neutrino and gamma-ray astronomy and the search for search for dark matter. Finally, the experimental prospects for the next decade in the 
VHE gamma-ray field are summarised.
\end{abstract}

\section{Introduction}

\begin{figure}[ht!]
\includegraphics[width=\textwidth,trim=45 40 45 95,clip]{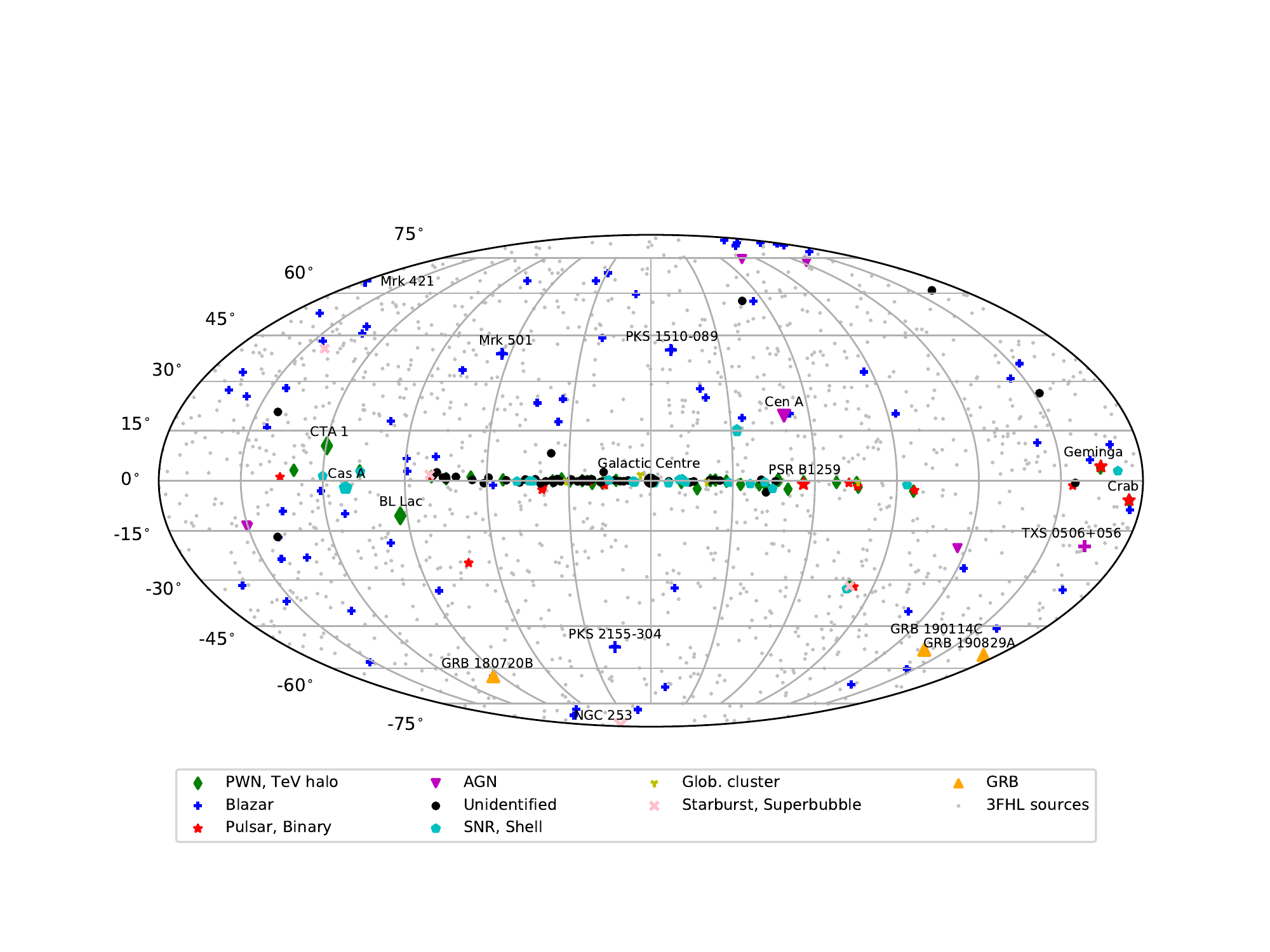}
\caption{\label{fig_sky}
The TeV sky in mid-2019. A compilation of known VHE gamma-ray sources (from \emph{TeVCat}), compared to the high energy Fermi-LAT catalogue (3FHL) sources. Adapted from~\cite{CTASciSympSum}.
}
\end{figure}

Astrophysical gamma-rays are a probe of cosmic particle acceleration and propagation, as well as a tool to probe beyond standard model (BSM) physics. Gamma-ray astronomy from the ground, rather than using satellites, is possible in the Very High Energy (VHE) domain above a few tens of GeV up to hundreds of TeV. This energy range is important for a number of reasons: it probes acceleration to the highest energies, provides access to the most promising range for BSM physics, and allows the highest angular resolution obtainable in any energy band above the X-ray range. Figure~\ref{fig_sky} illustrates the current knowledge of VHE gamma-ray sources. The number of known objects in this band ($\sim$200) is not impressive, but there is a huge variety to these objects, indicating that acceleration of particles to TeV energies is a rather common astrophysical process. Strong links exist between VHE gamma-ray astronomy and the neighbouring fields of cosmic particle detection, TeV-PeV neutrino astronomy and to gravitational wave astronomy. Here we provide a few examples of hot topics in multi-messenger astrophysics in which VHE gamma-ray observations play a key role. Given the recent developments in the field, we focus in particular on the area of explosive transients.

\section{The Gamma-ray/Neutrino Connection}

Interactions of ultra-high energy protons and nuclei result in the production of pions and hence neutrino and gamma-rays fluxes. As a consequence both neutrinos and gamma-rays have long been recognised as powerful probes of the cosmic rays in our own galaxy and beyond. The complementary of the two approaches
relies on the unambiguous nature of the neutrino emission (versus the potential for confusion in the VHE gamma) combined with the sensitivity and precision of gamma-ray telescopes (with respect to neutrino telescopes). Another important aspect is the transparency of the universe to neutrinos (in comparison to the gamma-ray horizon at moderate redshift at VHE), noting that this may in some cases make source identification harder (for example with respect to the ultra-high-energy cosmic rays, where the GZK horizon leads to a dramatic reduction in the number of candidate sources). 

Particularly in the case of transient events, the very large area of VHE gamma-ray instruments (in particular the Imaging Atmospheric Cherenkov Telescopes -- IACTs) makes them an ideal counterpart to neutrino telescopes, with typically thousands of detected gamma-rays expected for each neutrino.
Interactions of accelerated nuclei both with matter and radiation fields lead to simultaneous neutrino and gamma-ray production. In general neutrino and gamma-ray fluxes from the two process are comparable and in the absence of absorption effects it is straightforward to predict the expected neutrino spectrum from the observed gamma-ray spectrum (see e.g.~\cite{Kappes_2007}). In the case of photo-hadronic interactions however the necessary presence of strong radiation fields makes gamma-gamma interactions and cascading very likely. This situation breaks the simple relationship between gamma-ray and neutrino fluxes, but the combination remains very powerful as a diagnostic of the underlying physics and physical conditions in the emission region.

For the proton accelerators in our own galaxy the p-p channel is the most promising, and extensive surveys exist of the Galactic Plane in both, neutrinos and VHE gamma-rays. Unfortunately there are so far no firmly identified Galactic neutrino sources, but it is intriguing to note that one of the most promising regions from the recent IceCube search~\cite{IceCubeSearch}, is coincident with a source (MGRO\,J1908+06) now established by the HAWC collaboration to emit TeV photons to energies beyond 100~TeV~\cite{HAWC_highE}.

Beyond our galaxy, the large dataset from IceCube now places tight constraints on cosmic-ray acceleration and neutrino production in both gamma-ray bursts (GRBs,~\cite{Aartsen_2016}) and the population of gamma-ray emitting active galaxies know as blazars~\cite{IceCube_BlazarLimit}. Whilst there is so far no evidence for neutrino emission from active galaxies as a population, there is one very important candidate object which is discussed in detail below.

\subsection{TXS 0506+056}

On the $22^{\rm nd}$ of September 2017, a muon-neutrino induced signal in IceCube with a reconstructed energy of 290 TeV was sent out as an alert, prompting a global response at all wavelengths from radio to VHE gamma-ray~\cite{TXS0506_MM}. This neutrino was quickly associated to the flaring active galaxy TXS\,0506+056, with a detection at VHE from the MAGIC telescope. Although the association of the neutrino event with this galaxy is not overwhelmingly statistically significant, it is strengthened by the observation of an earlier period (between 2014 and 2015) when $13\pm5$ neutrino events in excess of the background expectation were seen from the direction of the source~\cite{IceCube_TXS0506_PreAlert}.

Several authors have pointed out the difficulty of resolving the measured gamma-ray and X-ray emission with the implied neutrino flux from TXS\,0506+056, in particular during the 2014/15 event (see e.g.~\cite{Rodrigues2019, Reimer2019}). In scenarios where the gamma-rays and neutrino emission are generated in the same zone by the same process, cascading results in gamma-ray emission that overshoots the observations. That the situation in TXS\,0506+056 is more complex than typical assumptions is hinted at by recent radio VLBI observations, which suggest
that this system contains two supermassive black holes producing jets that occasionally interact~\cite{Britzen2019}. This raises the possibility that proton acceleration in AGN jets may be very common, but neutrino flares from these objects very rare. 

\section{Explosive Transients} 

\subsection{Gravitational Wave Events}

Gravitational waves (GWs) originate from the coalescence of binary compact objects (black holes and neutron stars) and were first detected in 2015~\cite{PhysRevLett.116.061102}. In October 2017, GW~170817 was detected by the VIRGO and LIGO collaborations and 1.7 seconds later Fermi-GBM detected the electromagnetic counterpart of this event: the short-duration GRB~170817 and the associated optical transient SSS17a. 
This event was the first identification of an 
electromagnetic counterpart to a GW event. The detected signal was consistent with the merger of a binary neutron-star system and verified for the first time the progenitor hypothesis of short GRBs. 
In the VHE regime, observations were carried out using H.E.S.S. and HAWC~\cite{Abbott_2017}. The relatively large field of view (compared for example to typical optical and radio 
telescopes) of H.E.S.S. allowed for a fast coverage of  
the localisation region, with the SSS17a direction observed {\it before} the optical identification, 5 hours after the burst, 
thanks to an optimised coverage strategy. 
For HAWC, the event entered in the field of view $\sim$8 hours after the trigger. 
Although these observations resulted in non-detections, the extracted upper limits start to constrain very-high-energy emission models for such events.

\subsection{Gamma-ray bursts at very high energies, the breakthrough in 2018-2019}

Gamma-ray bursts (GRBs) are bright flashes of gamma rays, rapidly releasing as much as 10$^{49}$-10$^{53}$\,erg
of energy (assuming isotropic emission). The event duration T$_{90}$, the time in which 90\% of the emission
(typically measured in the 50 to 300\,keV energy range) is detected, of GRBs presents a bi-modal distribution.
Short GRBs (T$_{90}<$2\,s) originate from the merging of compact objects (see e.g.~\cite{Abbott_2017}, while the origin of long GRBs (T$_{90}>$2\,s) is the gravitational collapse of massive stars (see e.g.~\cite{Xu_2013}). 
The emission of GRBs comprises an initial episode (usually within the T$_{90}$
time window) referred as \emph{prompt} emission, and with an assumed \emph{internal} origin, 
followed by an extended (several hours or days) gradually-fading period known as \emph{afterglow}, which is thought to originate from the interaction of 
the ejected material with the circumburst medium 
(see~\cite{Meszaros2006, Nava2018} for a complete review).  

Whether or not GRBs are capable of emitting gamma-rays at VHE 
remained until recently an open question (see e.g.~\cite{Nava2018}). 
The strongest indication of VHE emission from GRBs was the detection of GRB~130427A (redshift z=0.34) using the \emph{Fermi} 
Large Area Telescope (LAT)~\cite{GRB130427A_LAT} including a 98 GeV 
photon detected at 243\,s after T$_0$ and a 32\,GeV photon detected 34\,ks 
after the onset. The energy of these photons significantly exceeded the 
synchrotron limit E$_{\rm syn,max}$ (obtained by equating the cooling to
the acceleration rate of particles in a decelerating blast wave,
see~\cite{Piran_2010} for details),
suggesting the rise of a new spectral component (most likely synchrotron
self-Compton) to account for the HE-detected 
emission~\cite{Liu_2013}.  Despite the expected synchrotron limit, evidence of 
a single spectral component up to the Fermi-LAT energy range was found, 
and a synchrotron origin remained as a feasible explanation 
for all of the detected emission~\cite{Nustar_2013}. GRB~130427A is one of 
the most remarkable afterglow observations thanks to its relatively nearby 
location, extremely bright emission and wide multi-wavelength coverage. 

\begin{figure}[h!]
    \centering
    \includegraphics[width = 0.85\textwidth]{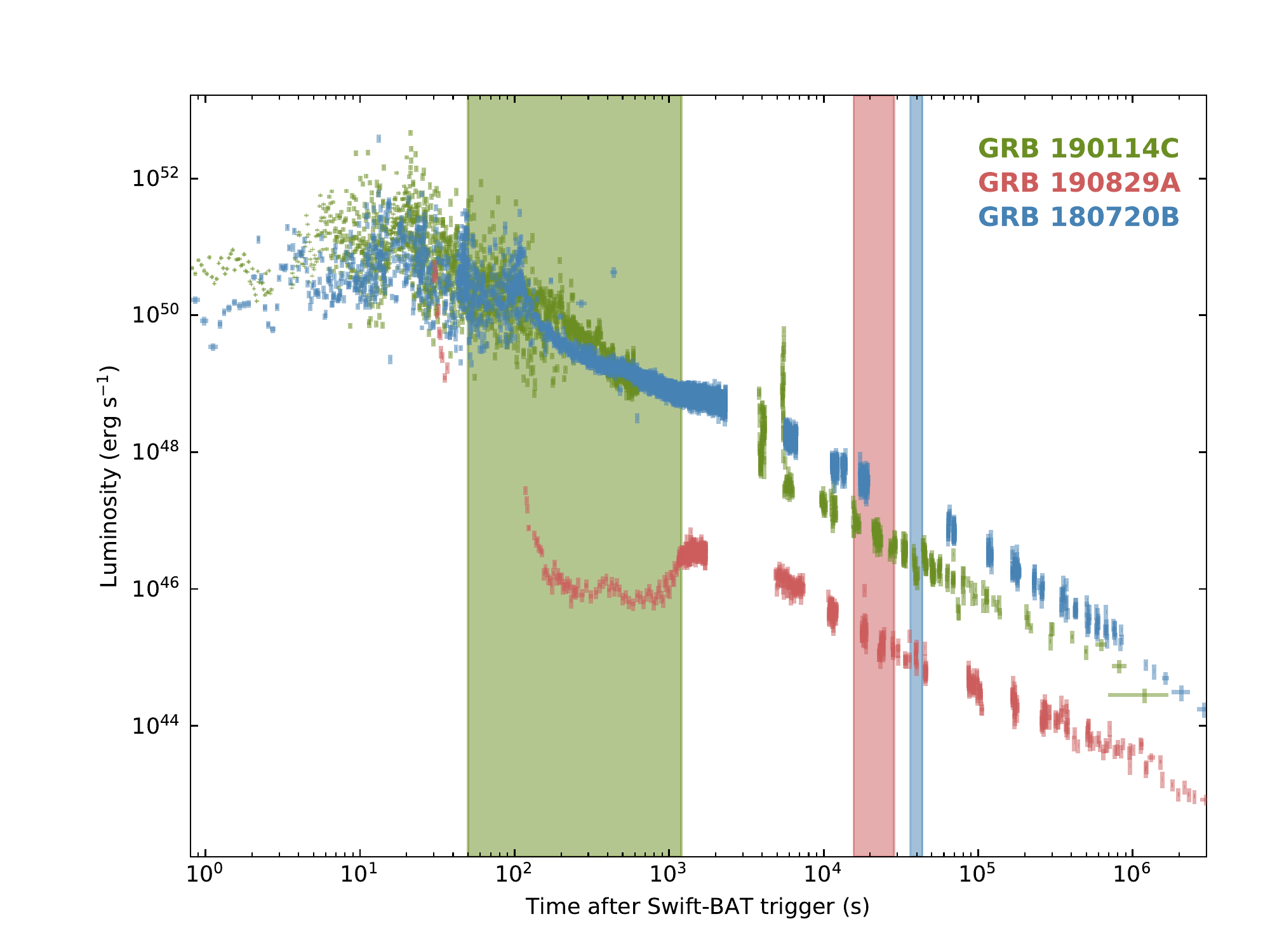}
    \caption{X-ray luminosity light curve of the three gamma-ray bursts with very-high-energy detection in 2018 and 2019. The shaded regions correspond to the observation windows of H.E.S.S. for GRB~180720B and GRB~190829A, and MAGIC for GRB~190114C. }
    \label{fig:threeGRBs}
\end{figure}

Since the origin of the detected HE emission of GRB~130427A remained under debate, VHE observations of GRBs
were strongly motivated. The measurement of VHE photons, together with  spectra at other wavelengths, 
can provide strong constraints on the circumburst medium characteristics, bulk Lorentz factor, 
and magnetic field strength, and provide insights into the emission and particle acceleration mechanisms at
relativistic shocks. Over many years there were extensive efforts towards VHE GRB detection
(see~\cite{HESSGRBsICRC2019,HAWCGRBsICRC2019,MAGICGRBsICRC2019,VERITASGRBs}). These programmes were finally,
and richly, rewarded in the last year with the afterglow detection of GRB~180720B with the H.E.S.S.
telescopes\footnote{GRB180720B TeVCat entry:\textit{http://tevcat.uchicago.edu/?mode=1;id=329} }, the
prompt-to-afterglow detection of GRB~190114C~\cite{ATelRazmik} with the MAGIC telescopes and once more with the
afterglow detection of GRB~190829A with H.E.S.S~\cite{AtelMathieu}. Figure~\ref{fig:threeGRBs} shows the X-ray luminosity light curve for each of these GRBs with the shaded region indicating the time coverage of each VHE detection and highlights the wide range of timescales in which VHE emission is being proved. 

The relatively small field of view of IACTs and the attenuation of VHE emission by pair-production on the extra-galactic background light are two of the main challenges to overcome for VHE observatories. These detections were possible due to the fast and precise localisation by satellites instruments (Swift-BAT, Fermi-GBM and LAT), the ambitious observations during moonlight periods or after long time delays, rapid telescope repointing, and finally the increased sensitivity of these instruments compared to previous VHE observatories and to satellite instruments such as Fermi-LAT at high energies. 

GRBs are key objects for multi-messenger studies for a number of reasons. For example, GRBs are one of the long-standing candidate sources of UHECRs, with the power required to match the UHECR flux being comparable to the total power output of GRBs in gamma-rays~\cite{Waxman2004}. However, searches for coincident high-energy neutrinos by the IceCube collaboration have shown no evidence for this association~\cite{Aartsen_2016}, placing strong constraints on UHECRs production models.
Other possibilities include low-luminosity GRBs as UHECRs factories~\cite{Zhang:2017moz}, although the detection of this type of GRB is currently severely limited by the sensitivity of present instruments. 

Successful observations such as those achieved in 2018 and 2019 are necessary to completely understand the acceleration mechanisms and emission processes of these events. These detections prove that GRB environments are capable of producing VHE radiation up to very late times, and have changed the paradigm and observational strategies for ground-based gamma-ray astronomy. Instruments with higher sensitivity such as CTA and later SWGO (see Sec.~\ref{sec:nextgen}) will allow for deeper and more detailed characterisation of these events over a wider range of timescales. 

So far, the three recent detections of GRBs at VHE correspond to the afterglow phase and are characterised as long GRBs. Although the environments and progenitors of the two classes of GRB are different, the X-ray afterglow of both types exhibit similar characteristics (temporal decay and spectral index), with short GRBs presenting only overall lower flux levels~\cite{Nysewander_2009}. VHE observations in the early-emission epoch of these events have a strong potential to extend the energetic and multi-messenger coverage and connection of GWs to the VHE gamma-ray regime.

\section{The search for Dark Matter}

Despite the so far negative results of searches for evidence of supersymmetry at the LHC, weakly interacting massive particles (WIMPs) remain a compelling candidate for Dark Matter~\cite{WIMP_Review}. In fact the non-detection LHC has increased interest in heavy WIMPs, in to the 0.1--10 TeV range that can be effectively probed by gamma-ray observations via WIMP annihilation as well as direct DM searches deep underground and anti-particle measurements with AMS. In the natural case of the WIMP as a thermal relic of the big bang, the velocity weighted cross section for self-annihilation is effectively fixed by the current DM density. This fact provides a clear target in terms of the required sensitivity of gamma-ray detectors, and this sensitivity level will be reached over a wide mass range by the next generation of detectors, as long as the universe is not too unkind in terms of the dark matter halo profile in our galaxy. 

In fact, for more favourable assumptions on both the halo profile and annihilation channel in to standard model particles, VHE telescopes are already reaching in to the thermal relic phase space. In particularly the H.E.S.S. limits from deep searches around the Galactic Centre~\cite{HESS_DMGC} reach the required cross section of $\langle \sigma v \rangle \approx 3\times10^{-26}$ cm$^{3}$ s$^{-1}$ for the $\tau^{+}\tau^{-}$ channel, for WIMP masses around 1~TeV. These VHE searches strongly complement the results from Fermi-LAT, which have excluded a WIMP with this cross-section with a mass below $\sim$100~GeV~\cite{FermiDM}.

Detection of anti-particles at the Earth provides a different method of indirect detection of WIMPs. Scenarios in which the anti-proton and positron fluxes are dominated by WIMP annihilation are already tightly constrained by diffuse gamma-ray limits~(see e.g.~\cite{Gaskins2016} and references therein). 
Advantages of gamma-ray detection of an annihilation signature over particle detection include the clear directional signature and the lack of assumptions necessary on cosmic ray transport within the galaxy.

As discussed below, a new generation of gamma-ray instruments is on its way and will have profound impact in the dark matter arena~(see e.g.~\cite{CTAScienceCase,Viana2019}).

\section{The next generation of VHE gamma-ray observatories}
\label{sec:nextgen}
\begin{figure}[ht!]
    \centering
    \includegraphics[width = 0.75\textwidth]{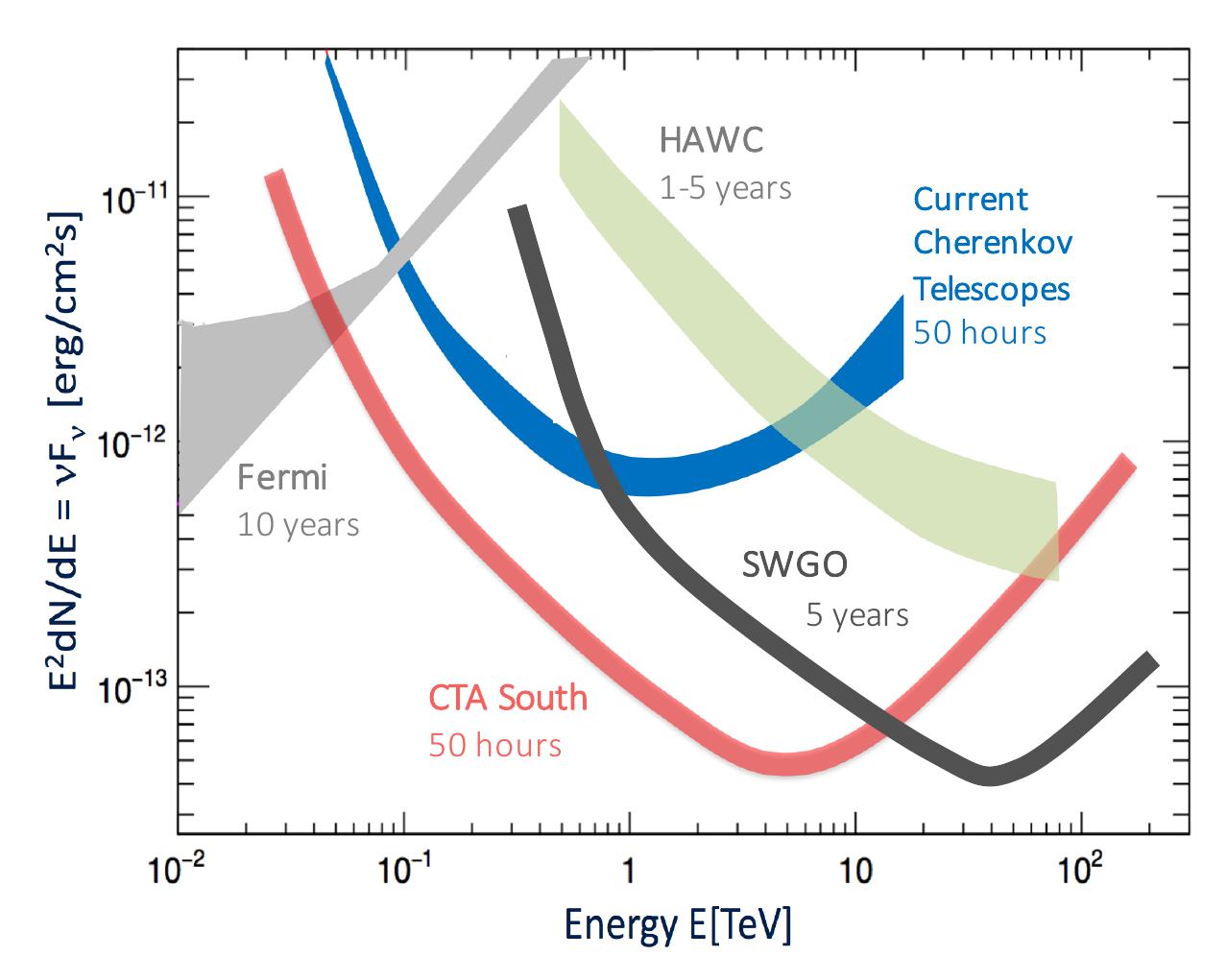}
    \caption{Comparison of the differential flux sensitivity of current and future VHE gamma-ray observatories. Adapted from~\cite{NIMWernerJim}, adding SWGO sensitivity from~\cite{SGSOsciencecase}.}
    \label{fig:sensitivity}
\end{figure}

The development of instrumental capabilities in the field of VHE gamma-ray astronomy is currently very active. Figure~\ref{fig:sensitivity} compares the sensitivity of a range of current and future instruments. One example of a recent upgrade is the extension of the HAWC array with a spare array of \emph{outriggers}~\cite{HAWCoutriggers}. This upgrade increases the array footprint by a factor of $\sim$4 in area and provides and improved precision and/or greater collection area for gamma rays with energies above $\sim$10\,TeV. 

One major new VHE gamma-ray observatory is under commissioning: the LHAASO project in Sichuan, China~\cite{LHAASO}. LHAASO contains a range of different detector types to cover shower energies from 100\,GeV to 1\,PeV and perform cosmic ray as well as gamma-ray science. From the perspective of VHE gamma-ray astronomy the most important component of LHAASO is the central water Cherenkov detector, which has an area a factor of four larger than the main array of HAWC. The addition of LHAASO means that the northern sky has excellent coverage in terms of wide-field high-duty cycle instrumentation. In the southern hemisphere no instrument of this type exists and there are many strong motivations for such a detector, including the search for DM annihilation in the Galactic Centre and emission from the central molecular zone and Fermi bubbles, as well as to complete the global coverage for transient and time-variable phenomena. The Southern Wide-field Gamma-ray Observatory (SWGO) collaboration recently formed to develop the plans for such an observatory, aiming for similar performance to that of LHAASO for VHE gammas~\cite{SWGO}.

Last, but not least, the primary gamma-ray facility of the next decades will be the Cherenkov Telescope Array (CTA)~\cite{CTAScienceCase}. CTA will have unprecedented precision and sensitivity across the VHE domain. Key science projects include a survey of the Galactic Plane and part of the extragalactic sky. CTA can provide precision measurements of transients or steady sources identified using LHAASO or SWGO, as well as targets identified at other wavelengths or in other messengers. The key role of VHE gammas in multi-messenger astronomy looks set to continue in the decades to come.


\bibliography{bib}{}

\end{document}